\documentclass[%
 amsmath,amssymb,
 aps,
]{revtex4-1}

\usepackage{graphicx}% Include figure files
\usepackage{dcolumn}% Align table columns on decimal point
\usepackage{bm}% bold math

\begin{document}

\title{Topological Order of Quantum Gravity in $AdS_3$ Spacetime}
\author{Jingbo Wang}
\email{ shuijing@mail.bnu.edu.cn}
\affiliation{Institute for Gravitation and Astrophysics, College of Physics and Electronic Engineering, Xinyang Normal University, Xinyang, 464000, P. R. China}
 \date{\today}
\begin{abstract}
Topological order is a new type order that beyond Landau's symmetry breaking theory. It has some interesting properties, such as producing quasi-particles with fractional quantum numbers and fractional/Fermi statistics, robust gapless boundary modes and emergent gauge excitations. In this essay, we will show that the quantum gravity in $AdS_3$ spacetime can also have topological orders. Actually the theory has all the three features that define the topological order. We conjecture that quantum gravity in four dimension can also have topological orders.
\end{abstract}
\pacs{04.70.Dy,04.60.Pp}
 \keywords{filling fraction; zero modes; fractional charges}
\bibliographystyle{unsrt}
\maketitle
\section{Introduction}
In our world there are many different phases of matter. For a long time, it was believed that those phases can be described by Laudau's symmetry breaking theory. But in late 1980s, it was found that in chiral spin liquid there may exist a new kind of order, so-called topological order that beyond the usual symmetry description \cite{to1}. The term ``topological" was motivated by the fact that the low energy effective theory for the chiral spin liquid is a topological quantum field theory.  Since then, the study of topological phases of quantum matter slowly becomes more and more active, and then a main stream in condensed matter physics now. It was proposed that such phase of matter can be used to build powerful quantum computer \cite{qc1,qc2}.

Topological order can be probed/defined by three essential properties \cite{to2}:
\begin{enumerate}
  \item  the ground state degeneracy on torus (or other space with non-trivial topology);
  \item  the non-Abelian geometrical phase of those degenerate ground states (which form a representation of the modular group $SL(2,Z)$);
  \item  the gapless edge modes.
\end{enumerate}
Those properties are all robust against any small perturbations. It can also be defined microscopically as pattern of long range entanglement. Topological order can produce quasiparticles with fractional quantum numbers and fractional/Fermi statistics, robust gapless boundary modes and emergent gauge excitations. So they can provide an unification of gauge interaction and Fermi statistics \cite{wen3}. A typical system which has topological order is the fractional quantum Hall states.

In this essay we will show that the quantum gravity in $AdS_3$ spacetime have topological orders.
\section{Topological order in 3D AdS gravity}
As first shown in Ref.\cite{at1}, $(2+1)-$dimensional general relativity can be written as a Chern-Simons theory. For $AdS_3$ spacetime which has negative cosmological constant $\Lambda=-1/L^2$, one can define two $SL(2,R)$ connection 1-forms
\begin{equation}\label{1}
    A^{(\pm)a}=\omega^a\pm \frac{1}{L} e^a,
\end{equation}where $e^a$ and $\omega^a$ are the orthonormal co-triad and spin connection 1-form, respectively, and $a=0,1,2$ is gauge group index. Then, up to a boundary term, the first order action of gravity can be rewritten as
\begin{equation}\label{2}\begin{split}
    I_{GR}[e,\omega]=\frac{1}{8\pi G}\int e^a \wedge ({\rm d}\omega_a+\frac{1}{2}\epsilon_{abc}\omega^b \wedge \omega^c)-\frac{1}{6L^2}\epsilon_{abc}e^a\wedge e^b \wedge e^c\\=I_{CS}[A^{(+)}]-I_{CS}[A^{(-)}],
\end{split}\end{equation}
where $A^{(\pm)}=A^{(\pm)a}T_a$ are $SL(2,R)$ gauge potential, $T_a$ are generators of $SL(2,R)$ group, and the Chern-Simons action is
\begin{equation}\label{3}
    I_{CS}[A]=\frac{k}{4\pi}\int {\rm Tr}\{A\wedge {\rm d}A+\frac{2}{3}A\wedge A \wedge A\},
\end{equation}
with
\begin{equation}\label{4}
    k=\frac{L}{4G \hbar}.
\end{equation}
The field equation is
\begin{equation}\label{5}
    F^{(\pm)}={\rm d}A^{(\pm)}+A^{(\pm)} \wedge A^{(\pm)}=0,
\end{equation}
implying that $A$ is a flat connection. Such a connection is completely determined by its holonomies around closed noncontractible curves $\gamma$.

In this essay we will show that the quantum gravity in $AdS_3$ spacetime have topological orders.

\begin{enumerate}
  \item Firstly let's consider the first condition, that is the ground state degeneracy on torus $T^2$. The torus has two non-contractable cycles $\gamma_1$ and $\gamma_2$, which satisfy $\gamma_1 \gamma_2=\gamma_2 \gamma_1$. The phase space can be described in terms of six gauge-invariant traces of $SL(2,R)$ holonomies $T_1^{\pm},T_2^{\pm}$ and $T_{12}^{\pm}$, which satisfy the non-linear Poisson bracket algebra \cite{np1}
\begin{equation}\label{6}
  \{ T_1^{\pm}, T_2^{\pm}\}=\mp \frac{1}{4 L}(T_{12}^{\pm}-T_1^{\pm}T_2^{\pm}),
\end{equation}
and cyclical permutations of $T_i^{\pm}, i=1,2,12$. The six traces $T_i^{\pm}$ provide an overcomplete description of spacetime geometry, since one has the Mandelstam identities. The traces can be represented classically as
\begin{equation}\label{7}
  T_1^{\pm}=\cosh \frac{r_1^{\pm}}{2},\quad  T_2^{\pm}=\cosh \frac{r_2^{\pm}}{2},\quad  T_{12}^{\pm}=\cosh \frac{r_1^{\pm}+r_2^{\pm}}{2},
\end{equation}
where the global parameters $r_i^{\pm}$ satisfy
\begin{equation}\label{8}
 \{ r_1^{\pm}, r_2^{\pm}\}=\mp \frac{1}{L},  \quad \{ r_a^+, r_b^-\}=0.
\end{equation}

Quantisation of (\ref{8}) gives
\begin{equation}\label{9}
 [ \hat{r}_1^{\pm}, \hat{r}_2^{\pm}]=\mp \frac{i \hbar}{L}.
\end{equation}

One can quantize the traces $T_i^{\pm}$ and also their algebra (\ref{6}). But one can also regard the quantized traces $\hat{T}_i^{\pm}$ as traces of diagonal operator-valued holonomy matrices $\hat{T}_i^{\pm}=\frac{1}{2}tr \hat{U}_i^{\pm},i=1,2$, and $\hat{T}_{12}^{\pm}=\frac{1}{2}tr (\hat{U}_1^{\pm}\hat{U}_2^{\pm})$, where the matrices $\hat{U}_i^{\pm}$ have the form
\begin{equation}\label{10}
 \hat{U}_i^{\pm}=\left(
             \begin{array}{cc}
               e^{\frac{\hat{r}_i^{\pm}}{2}} & 0 \\
               0 & e^{-\frac{\hat{r}_i^{\pm}}{2}} \\
             \end{array}
           \right)=\exp(\frac{ \hat{r}_i^{\pm}\sigma_3}{2}).
\end{equation}
From the fundamental relation (\ref{9}) one can get
\begin{equation}\label{11}
  \hat{U}_1^{\pm} \hat{U}_2^{\pm}=e^{\mp \frac{i \hbar}{4 L}}\hat{U}_2^{\pm} \hat{U}_1^{\pm}=e^{\mp \frac{2 \pi i }{2 k}}\hat{U}_2^{\pm} \hat{U}_1^{\pm},
\end{equation}
that is, a deformation of the classical equation stating that the holonomies commute. The similar algebra also appeared in Ref.\cite{imb1}. This algebra can't be realized on a single ground state. This immediately tell us that the ground state must be degenerate, and the smallest representation has dimension $2k$. This is the same as the fractional quantum Hall states with filling factor $v=\frac{1}{2k}$. So we show that the ground states on a torus are indeed degenerate, which satisfy the first condition of the topological order.
  \item Next we consider the action of modular group on those degenerate ground states. It is well known that general relativity is invariant under spacetime diffeomorphism.  But if the spacetime manifold $M$ is topologically nontrivial, its group of diffeomorphisms may not be connected: $M$ may admit “large” diffeomorphisms, which cannot be built up smoothly from infinitesimal deformations. The group of such large diffeomorphisms, $D(M)$, is called the mapping class group of $M$; for the torus $T^2$, it is also known as the modular group $SL(2,Z)$. Classically, geometries that differ by actions of $D(M)$ are exactly equivalent. But for quantum gravity, it is not clear to consider the $D(M)$ as pure gauge or a real symmetry, see for example Ref.\cite{peld1,carl1}. If we choose the second opinion, that is, the wavefunctions transform just covariant under $D(M)$, the second condition for topological order is satisfied. For $SL(2,R)$ Chern-Simons theory with level $k \in N$, the modular group $SL(2,Z)$ can be represented on the Hilbert space with \cite{imb1}
\begin{equation}\label{12}\begin{split}
    (T(2k))_{mn}=\delta_{mn} \exp{(-2\pi i \frac{\theta(1;2k)}{3})}\exp{(\frac{i\pi}{2k}m^2)},\\
  (S(2k))_{mn}=\frac{1}{\sqrt{2k}} \exp{(\frac{2\pi i}{2k}mn)},
\end{split}\end{equation}
where the phase $\theta(1;2k)$ is determined by the $SL(2,Z)$ relation $(ST)^3=1$ and can be written as a Gauss sum
\begin{equation}\label{12a}
  \exp{(2\pi i \theta(1;2k))}=\frac{1}{\sqrt{2k}} \sum_{n=0}^{2k-1} \exp{(\frac{i\pi}{2k}n^2)}.
\end{equation}

On the other hand, the representation of the modular group on the fractional quantum Hall states with filling factor $v=\frac{1}{2k}$ is given by \cite{modular1,modular3}
\begin{equation}\label{13}\begin{split}
  (T(2k))_{mn}=\delta_{mn} \exp{(-2\pi i \frac{c}{24})} \exp{(\frac{i\pi}{2k}m^2)},\\
  (S(2k))_{mn}=\frac{1}{\sqrt{2k}} \exp{(\frac{2\pi i}{2k}mn)},
\end{split}\end{equation}
where $c$ is the central charge. They form a unitary representation of the modular group $SL(2,Z)$. The (\ref{12}) and (\ref{13}) is the same if we identity $c=8\theta(1;2k)$.

  \item Finally we consider the edge modes. We can consider the horizon of BTZ black hole as the boundary of spacetime. As was shown in Ref.\cite{whcft1}, those boundary degrees of freedom can be described by two chiral massless scalar fields $\Psi_L$ (for left-moving) and $\Psi_R$ (for right-moving). From the analysis of the quasi-particles, it was shown that the BTZ black holes can be consider as the fractional quantum Hall states with filling factor $v=\frac{1}{2k}$ \cite{wangti3}. So the third condition is also satisfied.
\end{enumerate}

In conclusion, the quantum gravity in $AdS_3$ Spacetime satisfies all the three conditions for topological order, so it has topological order. And it seems to have the same topological order with the quantum spin Hall state which can be described by the Chern-Simons theory with the $K-$matrice $K=\left(
                                                                                                                           \begin{array}{cc}
                                                                                                                             2k & 0 \\
                                                                                                                             0 & -2k \\
                                                                                                                           \end{array}
                                                                                                                         \right)$, since they have the same modular data $(S,T)$.

In four dimensional spacetime, the black holes also have massless boundary modes \cite{wangbms4}, so the third condition is still satisfied. So we conjecture that the quantum gravity in  four dimensional spacetime also have topological order.
\acknowledgments
 This work is supported by Nanhu Scholars Program for Young Scholars of XYNU.

%\bibliography{to1}

\begin{thebibliography}{}

\end{thebibliography}


\begin{thebibliography}{10}

\bibitem{to1}
X.~G. {Wen}.
\newblock {Topological Orders in Rigid States}.
\newblock {\em International Journal of Modern Physics B}, 4(2):239--271,
  January 1990.

\bibitem{qc1}
A.~Y. Kitaev.
\newblock {Fault tolerant quantum computation by anyons}.
\newblock {\em Annals Phys.}, 303:2--30, 2003.

\bibitem{qc2}
C.~Nayak, S.~H. Simon, A.~Stern, M.~Freedman, and S.~Das~S.
\newblock {Non-Abelian anyons and topological quantum computation}.
\newblock {\em Rev. Mod. Phys.}, 80:1083--1159, 2008.

\bibitem{to2}
X.~G. Wen.
\newblock {Choreographed entangle dances: topological states of quantum
  matter}.
\newblock {\em Science}, 363(6429), 6 2019.

\bibitem{wen3}
B.~Zeng, X.~Chen, D.-L. Zhou, and X.~G. Wen.
\newblock {\em {Quantum Information Meets Quantum Matter: From Quantum
  Entanglement to Topological Phases of Many-Body Systems}}.
\newblock Springer, New York, 2019.

\bibitem{at1}
A.~Achucarro and P.~K. Townsend.
\newblock {A Chern-Simons Action for Three-Dimensional anti-De Sitter
  Supergravity Theories}.
\newblock {\em Phys. Lett.}, B180:89, 1986.

\bibitem{np1}
J.~E. Nelson and R.~F. Picken.
\newblock {Quantum holonomies in (2+1)-dimensional gravity}.
\newblock {\em Phys. Lett. B}, 471:367--372, 2000.

\bibitem{imb1}
C.~Imbimbo.
\newblock {SL(2,R) Chern-Simons theories with rational charges and
  two-dimensional conformal field theories}.
\newblock {\em Nucl. Phys. B}, 384:484--506, 1992.

\bibitem{peld1}
P.~Peldan.
\newblock {Large diffeomorphisms in (2+1) quantum gravity on the torus}.
\newblock {\em Phys. Rev. D}, 53:3147--3155, 1996.

\bibitem{carl1}
S.~Carlip.
\newblock {Quantum gravity in 2+1 dimensions: The Case of a closed universe}.
\newblock {\em Living Rev. Rel.}, 8:1, 2005.

\bibitem{modular1}
X.~G. Wen.
\newblock Topological orders in rigid states.
\newblock {\em Int. J. Mod. Phys. B}, 4(2):239, 1990.

\bibitem{modular3}
X.~G. Wen.
\newblock {A theory of 2+1D bosonic topological orders}.
\newblock {\em Natl. Sci. Rev.}, 3(1):68--106, 2016.

\bibitem{whcft1}
J.~Wang and C.-G. Huang.
\newblock {Conformal field theory on the horizon of a BTZ black hole}.
\newblock {\em Chin. Phys.}, C42(12):123110, 2018.

\bibitem{wangti3}
J.~Wang.
\newblock {Classification of black holes in three dimensional spacetime by the
  $W_{1+\infty}$ symmetry}.
\newblock {\em ArXiv:1804.09438}, 2018.

\bibitem{wangbms4}
J.~Wang.
\newblock {Microscopic states of Kerr black holes from boundary-bulk
  correspondence}.
\newblock {\em Chin. Phys.}, C45(1):015107, 2021.

\end{thebibliography}

\end{document}